\documentclass[letterpaper, 10pt, conference]{ieeeconf}

\usepackage[top=1.9cm, left=1.9cm, right=1.9cm, bottom=1.9cm]{geometry}
\usepackage[T1]{fontenc}
\usepackage{amsmath,amssymb,amsfonts}
\usepackage{graphicx,epstopdf}
\usepackage{graphics}
\usepackage{url}           
\usepackage{hyperref}
\usepackage[normalem]{ulem}
\usepackage{algorithm}
\usepackage{algorithmic}
\usepackage{booktabs}
\usepackage{cite}
\usepackage{balance}

\graphicspath{ {.figures/} }
\IEEEoverridecommandlockouts                              

\DeclareUrlCommand{\ULurl}{%
	\renewcommand\UrlLeft{\uline\bgroup}%
	\renewcommand\UrlRight{\egroup}}

\def\BibTeX{{\rm B\kern-.05em{\sc i\kern-.025em b}\kern-.08em
		T\kern-.1667em\lower.7ex\hbox{E}\kern-.125emX}}

\expandafter\def\expandafter\normalsize\expandafter{%
    \normalsize%
    \setlength\abovedisplayskip{4.5pt}%
    \setlength\belowdisplayskip{6pt}%
    \setlength\abovedisplayshortskip{-8pt}%
    \setlength\belowdisplayshortskip{2pt}%
}

\renewcommand{\baselinestretch}{0.95} 

\begin{document}
    
    \title{Multiple Target Tracking Using a {UAV} Swarm in Maritime Environments}
    
    \author{Andreas~Anastasiou, Savvas~Papaioannou, Panayiotis~Kolios,~and~ Christos G.~Panayiotou
        \thanks{Authors are with KIOS Research and Innovation Center of Excellence (KIOS CoE). A. Anastasiou and C. Panayiotou are also with the Department of Electrical and Computer Engineering, and P. Kolios is with the Department of Computer Science, University of Cyprus. E-mail:\texttt{\{anastasiou.antreas, papaioannou.savvas, pkolios, christosp\}} @ucy.ac.cy}
        \thanks{This work is implemented under the Border Management and Visa Policy Instrument (BMVI) and is co-financed by the European Union and the Republic of Cyprus (GA:BMVI/2021-2022/SA/1.2.1/015), and supported by the European Union's Horizon 2020 research and innovation programme under grant agreement No 739551 (KIOS CoE), and through the Cyprus Deputy Ministry of Research, Innovation and Digital Policy of the Republic of Cyprus.}}
    \maketitle
    
    \begin{abstract}
        Nowadays, unmanned aerial vehicles (UAVs) are increasingly utilized in search and rescue missions, a trend driven by technological advancements, including enhancements in automation, avionics, and the reduced cost of electronics. In this work, we introduce a collaborative model predictive control (MPC) framework aimed at addressing the joint problem of guidance and state estimation for tracking multiple castaway targets with a fleet of autonomous UAV agents. We assume that each UAV agent is equipped with a camera sensor, which has a limited sensing range and is utilized for receiving noisy observations from multiple moving castaways adrift in maritime conditions. We derive a nonlinear mixed integer programming (NMIP) -based controller that facilitates the guidance of the UAVs by generating non-myopic trajectories within a receding planning horizon. These trajectories are designed to minimize the tracking error across multiple targets by directing the UAV fleet to locations expected to yield targets measurements, thereby minimizing the uncertainty of the estimated target states. Extensive simulation experiments validate the effectiveness of our proposed method in tracking multiple castaways in maritime environments.
    \end{abstract}
    \begin{keywords}
        Search and Rescue Robots,
        Model Predictive Control,
        Multi-agent Systems
    \end{keywords}
	
    \section{Introduction}\label{intro} 
    Tracking multiple targets in maritime environments is challenging due to vast operational areas and sparse observations, which cause positional estimation errors to accumulate. Recent advancements in unmanned aerial vehicle (UAV) technology have enabled their deployment in emergency operations such as search and rescue (SAR) \cite{papaioannou2021towards,papaioannou2020coordinated,anastasiou2020swarm}, search and track (SAT) \cite{anastasiou2021hyperion,anastasiou2023model}, and reconnaissance missions \cite{papaioannou2020cooperativeSecurity,papaioannou2021downing}, as well as in critical infrastructure inspection \cite{savva2021icarus,zacharia2023distributed,papaioannou2023integrated,anastasiou2024automated} and disaster management \cite{terzi2019swifters,papaioannou2024synergising}. Deploying multiple autonomous UAVs in maritime scenarios is vital for disaster management, security, and safety at sea. The number of lost ships and shipwrecks globally remains alarmingly high~\cite{allianz2023lost}, while urgent migration forces many to undertake dangerous journeys across hazardous terrains and seas, especially in Mediterranean countries. For instance, in 2023, over 270,000 people crossed the Mediterranean toward European nations such as Italy, Spain, Greece, and Cyprus, with more than 4,000 reported dead or missing according to the United Nations High Commissioner for Refugees~\cite{mediterranean2024migrants}.

    Maritime survivor tracking, known as the castaway problem, has received significant attention due to the diverse scenarios and conditions encountered in practice \cite{liu2023survey}. Various methods have been proposed for single and multiple target tracking using single and multiple agents. Feraru et al. \cite{feraru2020Towards} employed a modified leeway model based on wind and current data and introduced a zig-zag pattern algorithm for single-agent tracking. Ramírez et al. \cite{ramirez2011coordinated} utilized an Artificial Neural Network to predict castaway locations and implemented a controller to minimize the distance between agents and estimated positions. Alotaibi et al. \cite{alotaibi2019lsar} proposed a layered search and rescue algorithm to optimize operation time and maximize lives saved, while Papaioannou et al. \cite{papaioannou2019decentralized} introduced an augmented probability hypothesis density filter with decentralized cooperative strategies for target density propagation. Oliveira et al. \cite{oliveira2016moving} developed a Lyapunov-based nonlinear controller for tracking moving targets, and Song et al. \cite{song2018multi} applied standoff tracking techniques.
    Despite these advancements, maritime environments pose unique challenges that remain underexplored. Drift caused by waves often separates or unites castaways into groups, complicating tracking efforts. To address this, modified Stokes equations are used to simulate realistic castaway motion influenced by waves in deepwater conditions \cite{shen2001theoretical}.
    
    Unlike existing approaches that focus on myopic plans, this work generates non-myopic trajectories within a receding planning horizon to minimize uncertainty in estimated target states. It accounts for measurement errors due to UAV altitude changes, which affect object detection using camera sensors and computer vision algorithms, primarily convolutional neural networks (CNNs) trained on labeled image data of varying resolutions. A piecewise function models this error to minimize the estimation covariance of each target during the mission.

    This paper proposes a collaborative model predictive control (MPC) framework for tracking multiple castaway targets in maritime settings using autonomous UAVs. The MPC is formulated as a Nonlinear Mixed Integer Program (NMIP) to optimize UAV control inputs online, minimizing tracking error within a receding planning horizon. UAV trajectories are designed to reduce state estimation covariance by directing UAVs to locations expected to provide target measurements, thereby minimizing the posterior covariance of target states. The specific contributions of this work are:
    \begin{itemize}
        
        \item We propose a collaborative multi-UAV MPC framework to minimize tracking error, i.e., state uncertainty, of multiple castaways in maritime environments. The controller, formulated as an NMIP, guides the UAV fleet with non-myopic trajectories within a receding planning horizon.
        
        \item To address maritime tracking challenges like target dispersion and coalescence, our framework integrates a target clustering scheme into the controller for efficient UAV allocation and improved tracking coverage.

        \item We simulate castaway drift in rough seas using realistic motion dynamics based on a modified Stokes' drift equation, enhancing scenario realism.

    \end{itemize}

    The paper is structured as follows: Section \ref{sec:preliminaries} covers preliminary concepts, Section \ref{sec:proposedApproach} details the proposed approach, Section \ref{sec:simulationExperiments} presents simulation results, and Section \ref{sec:conclusions} concludes with future work directions.

    \section{Preliminaries}\label{sec:preliminaries}
        \subsection{System Architecture}
        \label{systemInteg}
\begin{figure}
                \centering
                \includegraphics[trim={0.1cm 0.95cm 0.1cm 0.7cm},clip,width=\columnwidth]{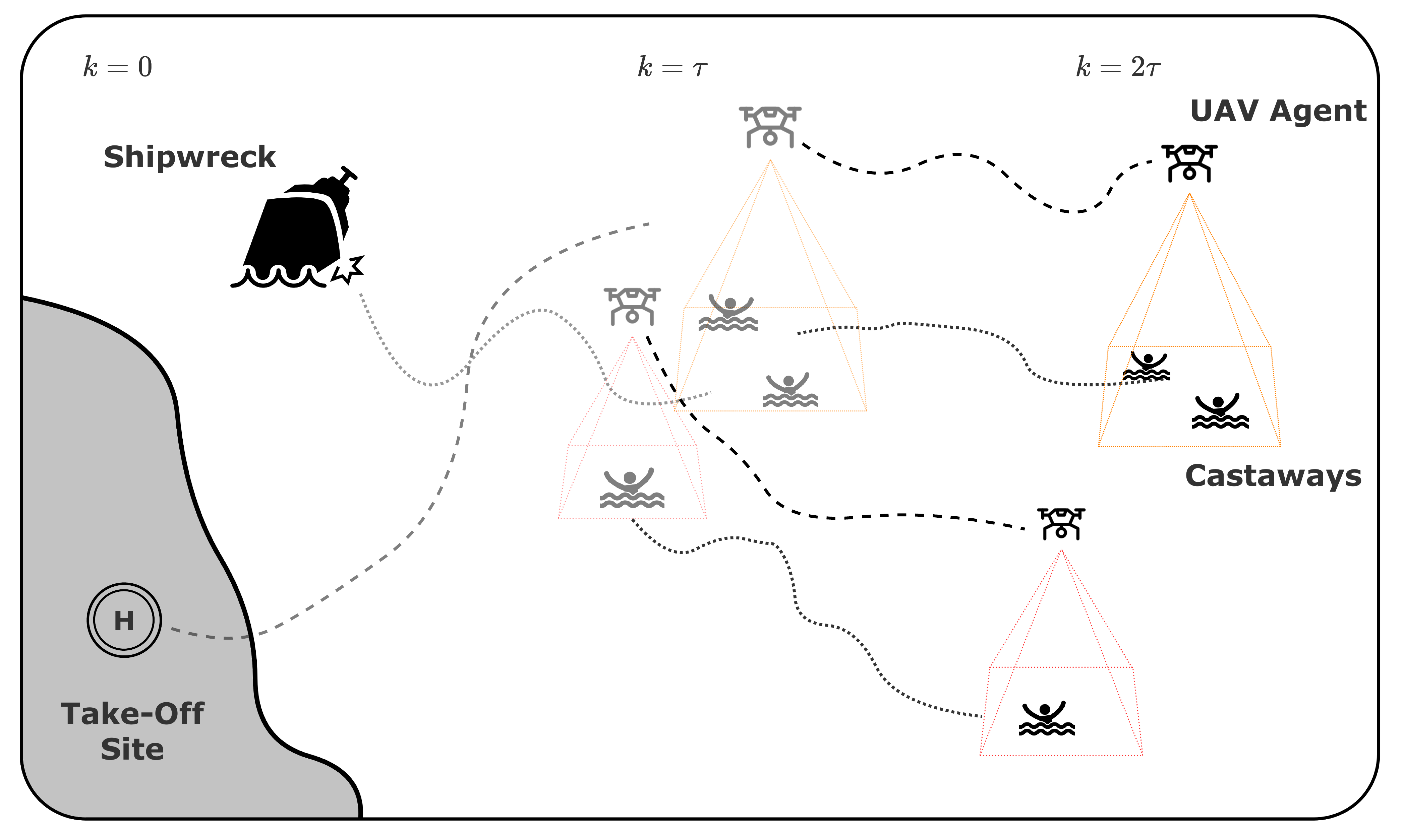}
                \vspace{-20pt}
                \caption{Scenario illustration of the proposed system.}
                \label{fig:scenarioPlot}
                \vspace{-15pt}
            \end{figure}

            The proposed architecture assumes a mission initiating system (e.g., coastal radar) that provides initial distributions of castaway states, including their number and positions, during a maritime incident. A fleet of $N$ UAV agents, each equipped with a camera sensor, observes the surroundings and continuously exchanges information to ensure all agents have the latest data. Using the initial distributions, each UAV initiates a Kalman Filter (KF) \cite{bishop2001introduction} for each castaway (targets). While castaways remain within the UAV's Field of View (FoV) (detailed in Sec.~\ref{ssec:sensing_model}), sensor measurements update KF predictions. If a castaway exits the FoV, the KF updates estimates without corrections, increasing positional uncertainty. To reduce uncertainty and improve location accuracy, UAVs must keep targets within their sensor range. Observing targets is probabilistic, governed by observation probability (Sec.~\ref{ssec:probDetection}), which decreases with UAV altitude, as illustrated in Fig.~\ref{fig:scenarioPlot}.

	\subsection{Agent Dynamics}
	\label{ssec:agent_dynamics}
        In this work, each UAV agent $i\in {1,\ldots,N}$ maneuvers in 3D Cartesian space, with its motion described by the discrete, linear dynamical model:
        \begin{equation}
            \footnotesize
            \label{eq:agentTransition}
            \mathbf{x}^i_{\tau+1} = A\mathbf{x}^i_{\tau} + Bu^i_{\tau}
        \end{equation}
        where the agent's state at time step $\tau$ is $\mathbf{x}^i_\tau = [\mathrm{p}^i_\tau, \mathrm{v}^i_\tau]^T$, comprising position $\mathrm{p}^i_\tau=[x^i_\tau, y^i_\tau, z^i_\tau]^T \in \mathbb{R}^{3}$ and velocity $\mathrm{v}^i_\tau=[\dot{x}^i_\tau, \dot{y}^i_\tau, \dot{z}^i_\tau]^T \in \mathbb{R}^{3}$. The control input vector $u^i_{\tau} = [u^i_x, u^i_y, u^i_z]^T \in \mathbb{R}^{3}$ represents the force applied in each dimension. The system matrices $A \in \mathbb{R}^{6\times6}$ and $B \in \mathbb{R}^{6\times3}$ are defined as:
        \begin{gather}
            \footnotesize
            A =
            \begin{bmatrix}
            \mathbf{I}_{3\times3} & \delta t \mathbf{I}_{3\times3}\
            \mathbf{0}_{3\times3} & \rho \mathbf{I}_{3\times3}\
            \end{bmatrix},
            B =
            \begin{bmatrix}
            \mathbf{0}_{3\times3}\
            \gamma \mathbf{I}_{3\times3}\
            \end{bmatrix}
        \end{gather}
        Here, $\delta t$ is the sampling interval, $\rho\in$ models air resistance, and $\gamma = {\frac{\delta t}{m}}$ converts control force to acceleration, with $m$ as the agent's mass. $\mathbf{I}_{3\times3}$ and $\mathbf{0}_{3\times3}$ represent the identity and zero matrices, respectively.

        \subsection{Sensing Model}
        \label{ssec:sensing_model}
        As stated, each UAV agent $i$ is equipped with a gimbaled camera sensor pointing toward local gravity, capable of identifying castaways and determining their 2D $\left(x,y\right)$-coordinates. The sensor's effective sensing area is a rectangle projected onto the sea surface, centered at the agent's current $\left(x,y\right)$ position. The rectangle's dimensions depend on the horizontal ($\theta$) and vertical ($\phi$) FoV angles, computed as:
        \begin{equation}
        \footnotesize
        l^i_{h} = 2z^i_\tau\tan(\theta/2), \quad l^i_{v} = 2z^i_\tau\tan(\phi/2)
        \end{equation}
        If a target is within the FoV, the sensor provides a measurement represented by the random set $H^i_\tau$. This set may be empty with probability $1-p^i_\tau,~ p^i_\tau \in$, or contain a single element with probability $p^i_\tau$ (see Sec.~\ref{ssec:probDetection}).
        In the latter case, the measurement function that produces element $\mathrm{y}_{\tau}^{ij},~j\in \left\{1,\ldots,\mathcal{C}\right\}$ for target $j$ received by agent $i$ is:
        \begin{subequations}
            \footnotesize
            \begin{gather}
                \mathrm{y}_{\tau}^{ij} = h(\mathbf{p}^{j}_{\tau}) + w(p^i_{\tau})\\
                h(\mathbf{p}^{j}_{\tau}) = \mathbf{I}_{2\times2}\mathbf{p}^{j}_{\tau}, \quad w(p^i_{\tau}) \sim \mathcal{N}(\mathbf{0}_{2\times1},~{p^i_\tau}^{-1} \zeta ~\mathbf{I}_{2\times1})
            \end{gather}
        \end{subequations}
        
        Here, $w(p^i_{\tau})$ represents normally distributed measurement noise with zero mean and standard deviation ${p^i_\tau}^{-1} \zeta$, where $\zeta$ is a scaling parameter.

        \subsection{Probability of observing a target}
        \label{ssec:probDetection}
        This section explains how the observation probability $p^i_\tau$, modeled as a piecewise linear function, is derived. The probability $p^i_\tau$ emulates a real-world computer vision detector and is incorporated into the onboard camera's sensing model (see Sec.\ref{ssec:sensing_model}). It is defined as:
        \begin{gather}
            \footnotesize
            p^i_{\tau} = \left\{
            \begin{matrix}
                \vspace{2pt}
                &1 &\text{if } z^i_\tau\le\alpha_1\\
                \vspace{2pt}
                &p_{min} &\text{if } z^i_\tau\ge\alpha_2\\
                &\beta_1 z^i_\tau +\beta_2 &\text{otherwise}
            \end{matrix}\right.
        \end{gather}
        
        Here, $\alpha_1$ and $\alpha_2$ adjust the altitudes where the probability reaches its maximum and minimum, while $\beta_1$ and $\beta_2$ determine the linear gradient. These parameters must satisfy $p^i_\tau \in [0, 1]$ to comply with the first probability axiom. The modeling assumptions have been validated through experimental tests, as detailed in \cite{anastasiou2023mpc}.

        \subsection{Castaway Motion Model}
        \label{ssec:castMotion}
        Each castaway floats on the ocean's surface and moves in 3D space, governed by an adapted form of the Stokes' drift equations~\cite{shen2001theoretical}. These equations simulate the real positions of the castaways $\mathbf{p}^{j}_{\tau} = [x^{j}_\tau, y^{j}_\tau, z^{j}_\tau]^T, \forall~j\in{1,..,\mathcal{C}}$, as shown in \eqref{eq:castaway}. The water velocity experienced by a castaway is described in Eq. \eqref{eq:waveq}, while its position is updated as per Eq. \eqref{eq:castStateUpdate}:
        \begin{subequations}
            \footnotesize
            \label{eq:castaway}
            \begin{gather}
                v^{j}_{\tau} = {\frac{\omega \textit{h}}{2}} e^{wd^{j}} \sin{(qd^{j}-\omega \tau)} \label{eq:waveq}\\
                \mathbf{p}^{j}_{\tau+1} = \mathbf{p}^{j}_{\tau} + {v^{j}_{\tau}}
                \begin{bmatrix}
                    \cos(\varphi) & \sin(\varphi) & 1
                \end{bmatrix}^T \delta t \label{eq:castStateUpdate}
            \end{gather}
        \end{subequations}
        
\begin{figure}
            \centering
            \includegraphics[width=\columnwidth]{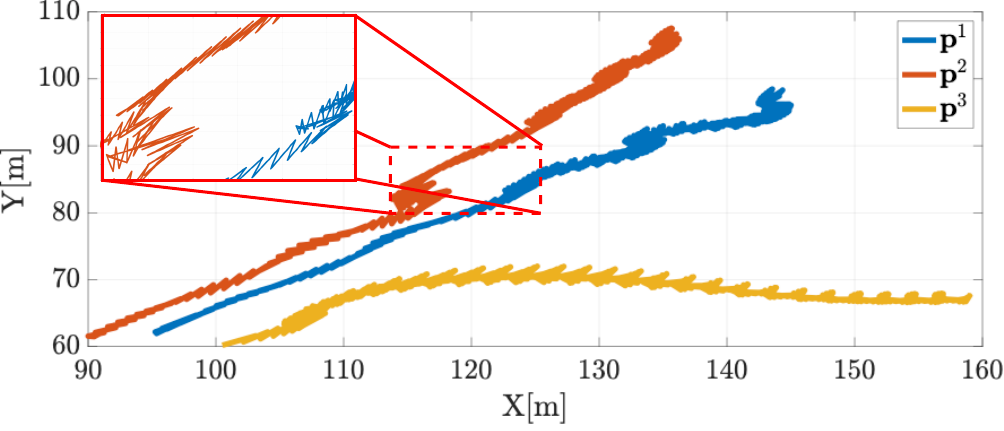}
            \vspace{-20pt}
            \caption{Drift paths of three castaways induced rough sea conditions over a time  span of 10 minutes.}
            \label{fig:scen1Plot}
            \vspace{-15pt}
        \end{figure}

        Here, $\omega={2\pi}/\mathcal{T}$ is the wave frequency, $h$ the wave height, $w$ the decay rate, and $d^{j}$ the distance of the $j^{th}$ castaway from the wave's origin. The wave number $q=2\pi/L$ depends on wavelength $L$, and $\varphi$ is the angle between the castaway's 2D location and the wave source. The wave period is $\mathcal{T}=\sqrt{2\pi L / gK}$, where $Z=\tanh(qD)$, $D$ is water depth, and $g$ is gravitational acceleration. Eq.~\eqref{eq:waveq} assumes small wave amplitude ($qh\ll1$), deep water conditions ($Z<1$), and no wave reflections. Fig.~\ref{fig:scen1Plot} illustrates three castaways drifting over 10 minutes.

    \section{Proposed Approach}
    \label{sec:proposedApproach}
        \begin{figure*}[t]
       		\centering
    		  \includegraphics[width=\textwidth]{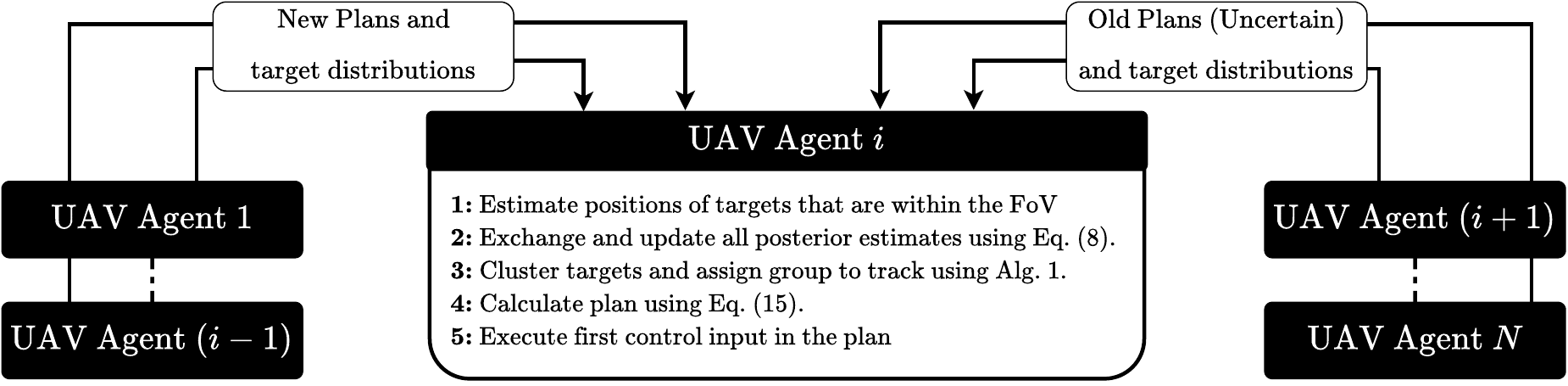}
                \vspace{-20pt}
    		  \caption{Flow diagram of the proposed approach showing the flow of exchanged information between agents and the sequence of steps that each agent follows.}
    		  \label{fig:flowDiagram}
    		  \vspace{-10pt}
        \end{figure*}
        This section details the proposed approach for tracking multiple castaways drifting after a maritime incident. Once initial castaway distributions are available, each UAV agent initializes a local KF state estimator for each castaway $j \in \{1,\ldots, \mathcal{C}\}$ to estimate and predict its location. Posterior state estimates are fused with those from other agents using the covariance intersection technique, which combines estimates when correlation information is unknown or unreliable.

        To optimize UAV fleet utilization and enhance tracking performance, a dynamic target clustering scheme is proposed. This scheme uses target state estimates and predicted paths to cluster targets into groups, anticipating future splitting scenarios that exceed a single UAV's tracking capacity. Targets are then allocated into separate groups, and each agent assigns itself a cluster in a greedy manner (see Sec. \ref{ssec:targetClustering}).

        A collaborative MPC framework (Sec. \ref{ssec:dmpc}) guides the $N$ UAV agents by generating non-myopic trajectories over a finite receding horizon $K$. These trajectories direct agents to locations yielding less noisy measurements, minimizing tracking error. Each agent plans sequentially, considering the fleet's current positions and future plans to ensure coordinated trajectories that minimize estimation covariance while avoiding collisions. The algorithm executed by each UAV agent is shown in Fig. \ref{fig:flowDiagram} and runs sequentially \cite{richards2007robust}, repeating until the mission ends.
                    
        \subsection{Target Clustering And Assignment}
        \label{ssec:targetClustering}
            Castaways tend to separate into groups during their drift, complicating multi-castaway tracking as agents must predict separations and assign castaways to groups based on drift direction and proximity to their previous group. To address this, we propose a deterministic clustering technique where each agent assigns targets into groups and selects a group to track based on proximity to the group's centroid. If all groups are already being followed, the agent tracks the group with the most castaways.

            Clustering begins with KF estimates of each target's future locations over the next $K$ time steps, enabling agents to predict drift direction. The clustering algorithm uses these estimates as input and ensures each target belongs to one group. It calculates each target's displacement in the $(x,y)$ plane and travel angle. Initially, all targets are assumed to belong to one group. Targets displaced beyond a threshold $d_g$ are assigned to new groups, and targets with similar travel angles in close proximity (Euclidean distance < $d_g$) are merged into the same group. Finally, agents greedily assign themselves to the group closest to their current position, based on the group's centroid distance.
            
        \subsection{Collaborative Model Predictive Control}
        \label{ssec:dmpc}
            The formulation in Eq. \eqref{eq:mpc} computes optimal control inputs for a UAV agent while considering neighboring agents. Each agent plans sequentially using the latest available plans from others in the fleet, as in \cite{richards2007robust}, enabling distributed tracking. Target clustering (Sec. \ref{ssec:targetClustering}) assigns each UAV $i \in \{1,\ldots,N\}$ to track $\mathcal{C}^a$ targets within its cluster. Agents share planned trajectories and target state distributions, enabling coordinated, non-myopic tracking and collision avoidance.
    
            Agents enhance estimates by exchanging target distributions and fusing them with neighbors' data using Eq. \eqref{eq:covariance_fusion}. Eq. \eqref{eq:covariance_fusion} (left) computes the weighted average of each target's state, while Eq. \eqref{eq:covariance_fusion} (right) calculates the weighted average of each target's covariance matrix. Weights $\textbf{W}_{i}$ minimize the covariance matrix trace \cite{julier2017general,Matzka2009}.
            \begin{equation}
                \footnotesize
                \label{eq:covariance_fusion}
                    \hat{\mathbf{p}}_{\tau \mid \tau}^{aj} =  \sum_{i=1}^N  \textbf{W}_{i}\hat{\mathbf{p}}_{\tau \mid \tau}^{ij}, \quad 
                    \hat{P}_{\tau \mid \tau}^{aj} = \sum_{i=1}^N \textbf{W}_{i} \hat{P}_{\tau+k+1 \mid \tau}^{ij} \textbf{W}^T_{i}
            \end{equation}
            The vector $u^i_{\tau}$, controlling UAV movement, is determined via a receding horizon MPC that adheres to the UAV's physical constraints, including velocity and acceleration limits. A NMIP problem is formulated to minimize the trace of the state covariance matrix $P^{j}_{\tau+k \mid \tau}$ for each predicted castaway position, as described in Sec. \ref{ssec:agent_dynamics}. KF equations are used to predict castaway drift, estimate positions, and quantify accumulated uncertainty over the planning horizon, especially when the target remains outside the agent's FoV for extended periods. Starting with imprecise measurements, the KF predicts future states and refines estimates using pseudomeasurements, noise approximations of expected sensor data when the castaway is within the FoV. The proposed MPC, represented as a NMIP in Eq. \eqref{eq:mpc}, calculates agent $a$'s plan.
            
            The objective, as outlined in Eq. \eqref{eq:mpcObj}, is to minimize the trace of the covariance matrix $P^{j}_{\tau+k \mid \tau}$ for every castaway throughout all time steps in the planning horizon.
            \begin{equation}
                \footnotesize
                \mathcal{J}^a =  \sum_{k=0}^{K-1}  \sum_{j=1}^{\mathcal{C}^a} tr(P^{j}_{\tau + k \mid \tau}) \label{eq:mpcObj}
            \end{equation}
            where, $K$ represents the total number of steps within the receding horizon, and $\mathcal{C}^a$ denotes the count of castaways included in the cluster being tracked by agent $a$. 
            \begin{align}
                &\mathbf{x}^a_{\tau+k+1 \mid \tau}=A \mathbf{x}^a_{\tau+k \mid \tau}+ Bu^a_{\tau+k \mid \tau}& \qquad \qquad \quad \forall~k \label{eq:agentMotion}
            \end{align}
            The UAV’s dynamic model, as mentioned earlier in Sec. \ref{ssec:agent_dynamics}, is depicted in Eq. \eqref{eq:agentMotion}. The UAV’s control input, denoted by the vector $u^i_{\tau+k \mid \tau}$, comprises $(u^i_x, u^i_y, u^i_z)$ for each step within the horizon $k\in \left\{0, \ldots, K-1\right\}$.
            \begin{subequations}
                \footnotesize
                \label{eq:FoVConstr}
                \begin{flalign}
                    &l^{i\kappa}_{\tau+k \mid \tau} = z^i_{\tau+k \mid \tau} \tan({\theta_\kappa}) & \forall~\kappa, k	\label{eq:fovSize}
                \end{flalign}
                \begin{flalign}
                    & \mathrm{d}^{i\kappa}_{\tau+k\mid \tau} = D_\kappa \mathbf{x}^i_{\tau+k\mid \tau} + (-1)^\kappa l^{i\kappa}_{\tau+k\mid \tau} &  \forall~i,\kappa, k\label{eq:fovLimits}\\
                    & b^{ij\kappa}_{\tau+k \mid \tau}=\left\{\begin{matrix}
                        1, &\text { if } E_\kappa \mathbf{p}^{j}_{\tau+k \mid \tau}\leq \mathrm{d}^{i\kappa}_{\tau+k \mid \tau} \\
                        0, &\text { otherwise }\\
                    \end{matrix}\right. &  \forall~i,j,\kappa,k\label{eq:binariesFOV}\\
                    & \mathrm{s}^{ij}_{\tau+k \mid \tau}=\sum_{\kappa} b^{{ij\kappa}}_{\tau+k \mid \tau} &  \forall~i,j, k \label{eq:binSum}\\
                    & \mathrm{b}^{ij}_{\tau+k \mid \tau}=\left\{\begin{matrix}
                        1, &\text { if } \mathrm{s}^{ij}_{\tau+k  \mid \tau} = 4 \\
                        0, &\text { otherwise }\\
                    \end{matrix}\right. & \forall~i,j, k \label{eq:binFoV}
                \end{flalign}
            \end{subequations}
            The constraints in Eq. \eqref{eq:FoVConstr} predict whether a target will fall within the agent's FoV. Eq. \eqref{eq:fovSize} calculates the horizontal and vertical dimensions of the camera's FoV, divided into halves $\kappa\in\{1,..,4\}$. Using these dimensions, Eq. \eqref{eq:fovLimits} estimates the FoV boundaries (left, right, top, bottom) in the 2D Cartesian plane. The matrix $D_\kappa \in \mathbb{R}^{6}$ selects the $(x,y)$ coordinates from the agent’s state vector.
            Constraints \eqref{eq:binariesFOV}, \eqref{eq:binSum}, and \eqref{eq:binFoV} determine a target's position relative to the FoV by checking if it lies within each boundary. Each check activates a binary variable $b^{ij\kappa}_{\tau+k \mid \tau}$ for target $j$ at time step $\tau$, using the big M method. The binary sum $\mathrm{s}^{ij}_{\tau+k \mid \tau}$ in Eq. \eqref{eq:binSum} indicates inclusion within the FoV. The matrix $E_\kappa \in \mathbb{R}^{3}$ selects $(x,y)$ coordinates from the target's state vector. If the sum equals 4, $\mathrm{b}^{ij}_{\tau+k \mid \tau}$ is activated (Eq. \eqref{eq:binFoV}), confirming that target $j$ is within agent $i$'s FoV at time step $\tau+k$.
            \begin{subequations}
                \footnotesize
                \label{eq:KalmanConstr}
                \begin{flalign}
                    & \mathbf{p}^{j}_{\tau+k+1 \mid \tau}=A^c \hat{\mathbf{p}}^{j}_{\tau+k \mid \tau} & \hspace*{-2.1mm} \forall~j, k \label{eq:priPos}\\
                    & P_{\tau+k+1 \mid \tau}^{j}=A^c \hat{P}_{\tau+k \mid \tau}^{j} {A^c}^{T}+Q & \hspace*{-2.1mm} \forall~j, k \label{eq:priCov}
                \end{flalign}
                \begin{flalign}
                    \begin{split}
                        K_{\tau+k+1 \mid \tau}^{ij}=&\hat{P}_{\tau+k+1 \mid \tau}^{j} C^{T} \\ \label{eq:gain}
                        &\left(C \hat{P}_{\tau+k+1 \mid \tau}^{j} C^{T}+R^i_{\tau+k \mid \tau}\right)^{-1}\\
                    \end{split}& \hspace*{-2.1mm} \forall~i,j, k\\
                    \begin{split}
                        \hat{\mathbf{p}}_{\tau+k+1 \mid \tau}^{\bar{i}j}=&\mathbf{p}_{\tau+k+1 \mid \tau}^{(\bar{i}-1)j}+\mathrm{b}_{\tau+k \mid \tau}^{\bar{i}j} K_{\tau+k+1 \mid \tau}^{\bar{i}j} \\ \label{eq:postPosRec}
                        & \left(\mathrm{y}_{\tau+k+1 \mid \tau}^{\bar{i}j}-C \hat{\mathbf{p}}_{\tau+k+1 \mid \tau}^{(\bar{i}-1)j}\right)\\
                    \end{split}& \hspace*{-2.1mm} \forall~\bar{i},j, k \\ 
                    \begin{split}
                        \hat{\mathbf{p}}_{\tau+k+1 \mid \tau}^{j}=&\mathbf{p}_{\tau+k+1 \mid \tau}^{Nj}+\mathrm{b}_{\tau+k \mid \tau}^{aj} K_{\tau+k+1 \mid \tau}^{aj} \\ \label{eq:postPos}
                        & \left(\mathrm{y}_{\tau+k+1 \mid \tau}^{aj}-C \hat{\mathbf{p}}_{\tau+k+1 \mid \tau}^{Nj}\right)\\
                    \end{split}& \hspace*{-2.1mm} \forall~j, k\\
                    \begin{split}
                        \hat{P}_{\tau+k+1 \mid \tau}^{\bar{i}j}=& P_{\tau+k+1 \mid \tau}^{(\bar{i}-1)j}-\mathrm{b}_{\tau+k \mid \tau}^{\bar{i}j} \\ \label{eq:postCovRec}
                        &\left(K_{\tau+k+1 \mid \tau}^{\bar{i}j} C P_{\tau+k+1 \mid \tau}^{(\bar{i}-1)j}\right)\\
                    \end{split}& \hspace*{-2.1mm} \forall~\bar{i},j, k 
                \end{flalign}
                \begin{flalign}
                    \begin{split}
                        \hat{P}_{\tau+k+1 \mid \tau}^{j}=& P_{\tau+k+1 \mid \tau}^{Nj}-\mathrm{b}_{\tau+k \mid \tau}^{aj} \\ \label{eq:postCov}
                        &\left(K_{\tau+k+1 \mid \tau}^{aj} C P_{\tau+k+1 \mid \tau}^{Nj}\right)\\
                    \end{split}& \hspace*{-2.1mm} \forall~j, k
                \end{flalign}
            \end{subequations}
            In addition, the constraints in \eqref{eq:priPos} to \eqref{eq:postCov} describe the KF estimator with the sequential update from the pseudomeasurements acquired by the rest of the agents in the fleet. The constraint \eqref{eq:priPos} concerns the estimate of the apriori state $\mathbf{p}^{j}_{\tau+k+1 \mid \tau} \in \mathbb{R}^4$ of target $j\in \mathcal{C}^a$ using the transition matrix $A^c=
            \begin{bmatrix}
                \mathbf{I}_{2\times2} & \delta t \mathbf{I}_{2\times2} \\ \mathbf{0}_{2\times2} & \mathbf{I}_{2\times2}
            \end{bmatrix}$.
            The identity and zero matrices are denoted by $\mathbf{I}_{2\times2}$ and $\mathbf{0}_{2\times2}$, and the sampling interval is $\delta t$. The KF is applied only to the target's $(x,y)$ positions in 2D, ignoring the Z-axis due to minor variations. Constraint \eqref{eq:priCov} computes the apriori covariance matrix, while constraint \eqref{eq:gain} calculates the KF gain for each UAV and target, requiring nonlinear constraints due to the inverse operation.
            Constraints \eqref{eq:postPosRec} and \eqref{eq:postPos} derive the posterior state estimation using pseudomeasurements for other agents $\bar{i}=\left\{1,\ldots,N\right\}-\left\{a\right\}$, while constraints \eqref{eq:postCovRec} and \eqref{eq:postCov} compute the posterior covariance matrix. Intermittent target observations are handled using the method from \cite{sinopoli2004kalman}, integrated into the NMIP framework with binary observation variables $\mathrm{b}^{ij}_{\tau+k \mid \tau}$ over the planning horizon.
            \begin{subequations}
                \footnotesize
                \label{eq:MeasConstr}
                \begin{flalign}
                    & \sigma^i_{\tau+k \mid \tau} = \lambda r(z^{i}_{\tau+k \mid \tau}) & \qquad  \forall~i,k \label{eq:measureNoise}\\
                    & R^i_{\tau+k\mid \tau}=\mathbf{I}_{2\times2}\sigma^i_{\tau+k \mid \tau} & \qquad  \forall~i,k \label{eq:measureCov}\\
                    & \mathrm{y}^{ij}_{\tau+k+1 \mid \tau} = C\hat{\mathbf{p}}_{\tau+k+1 \mid \tau}^{j} + n(\sigma^i_{\tau+k \mid \tau}) & \qquad \forall~i,j, k \label{eq:psudomeasure}
                \end{flalign}
            \end{subequations}            
            Constraint \eqref{eq:measureNoise} calculates the measurement's standard deviation using the tuning parameter $\lambda$ and the function $r(z^i_{\tau+k \mid \tau})$, which outputs a value in $[0,1]$ based on the agent's altitude, mimicking the piecewise function in \ref{ssec:probDetection}. As altitude increases, so does the standard deviation, and the observation noise covariance is computed accordingly in constraint \eqref{eq:measureCov}.
            During the planning horizon, pseudomeasurements are used to predict the target's future position and covariance matrix. Defined in constraint \eqref{eq:psudomeasure}, these pseudomeasurements are generated by adding zero-mean Gaussian noise $n(\sigma^i_{\tau+k \mid \tau}) \sim \mathcal{N}(0,\sigma^i_{\tau+k \mid \tau})$ to the apriori state estimate of the castaway's location. The observation matrix $C \in \mathbb{R}^{2\times3}$ extracts the $(x,y)$ coordinates from the target's state.
            \begin{align}
                d_{\tau+k \mid \tau}^{\bar{i}} = \lVert \mathbf{x}^{a}_{\tau+k \mid \tau} - \mathbf{x}^{\bar{i}}_{\tau+k \mid \tau} \rVert^2  & \qquad\qquad\qquad\quad \forall~\bar{i}, k \label{eq:colission}
            \end{align}
            To avoid inter-agent collisions, constraint \eqref{eq:colission} is employed to ensure that the Euclidean distance between agent $a$ and the rest of the agents is always at most $d_t$.

            \begin{subequations}
                \footnotesize
                \label{eq:mpc}
                \begin{align}
                    \hline \notag
                    \vspace{2pt}
                    &\qquad \qquad \qquad \min_{u^a_{\tau+k \mid \tau}} \quad \mathcal{J}^a & \notag \\
                    &\textbf{subject to} ~ i \in\left\{1,\dots,N\right\} \textbf{:}  & \notag\\
                    &\eqref{eq:agentMotion}-\eqref{eq:colission}&\\
                    & d_{\tau+k \mid \tau}^{\bar{i}} \geq d_{t}, \mathbf{x}^i_{\tau \mid \tau} \in \mathcal{X}, u^i_{\tau \mid \tau} \in \mathcal{U} & \enspace \quad \forall~\bar{i}, i \label{eq:boundaries} \\
                    & b^{ij\kappa}_{\tau+k \mid \tau}, \mathrm{b}_{\tau+k \mid \tau}^{ij} \in \left\{0,1\right\} & \enspace \quad \forall~i,j,\kappa, k \label{eq:binaryLims}\\
                    & \mathrm{s}^{ij}_{\tau+k \mid \tau} \in \{0,..,4\} & \enspace \quad \forall~i,j,\tau \label{eq:integerLims}\\
                    \hline \notag
                \end{align}
            \end{subequations}
            In conclusion, the NMIP can be represented concerning the previously stated constraints as shown in Eq. \eqref{eq:mpc}. Equations \eqref{eq:boundaries} set the safety distance limit, ensuring the UAV remains within the 3D space $\mathcal{X}$ and the input vector stays within the agent's practical capabilities $\mathcal{U}$. \vspace{3pt}

    \section{Simulation Experiments}
    \label{sec:simulationExperiments}
	
        \subsection{Simulation Setup} 
            \label{ssec:simulationSetup}
            As described in \ref{ssec:castMotion}, ground truth trajectories for castaways are generated to evaluate the framework via simulation. These trajectories account for drift after 10 minutes, influenced by wave sources with varying amplitudes $h$, decay rates $w$, and wavelengths $L$. The water depth $D$ was constant, satisfying small wave amplitude ($q\cdot h \ll 1$) and deepwater ($Z<1$) conditions. Fig. \ref{fig:scen1Plot} shows the trajectories, with each colored line representing a castaway's path.

            Each UAV's onboard camera generates measurements by adding zero-mean Gaussian noise to the castaways' ground truth, with noise variance proportional to the UAV's altitude. Data association was not considered, assuming agents can differentiate between targets.

            Realistic motion and input constraints were applied based on commercially available UAVs, limiting horizontal velocity to $\pm12m/s$, vertical velocity to $\pm7m/s$, and acceleration to $\pm7m/s^2$ in all dimensions. The FoV was set to $30^\circ$ horizontally and $20^\circ$ vertically, with the receding horizon limited to $K=3$ steps.
            
        \subsection{Simulation Results}
        \label{ssec:simulationResults}
            \begin{figure}
                \centering
                \includegraphics[width=\columnwidth]{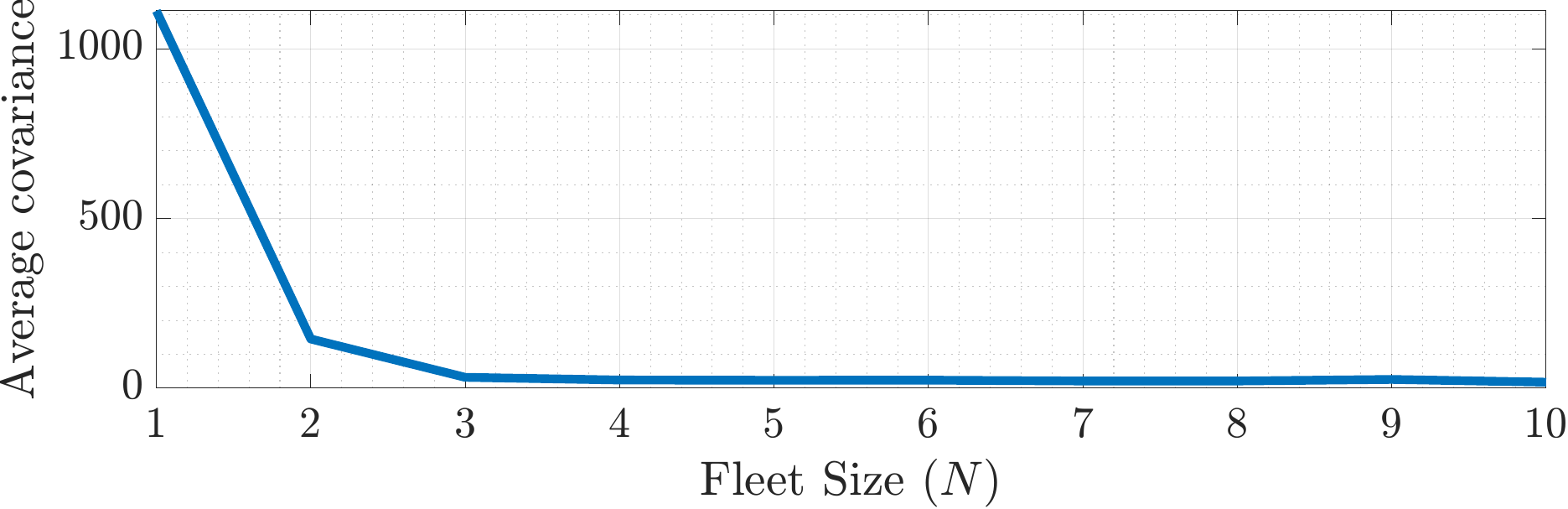}
                \vspace{-20pt}
                \caption{Average target estimation covariance based on fleet size.}
                \label{fig:covVSfleet}
                \vspace{-10pt}
            \end{figure}
            
            \begin{figure}
                \centering
                \includegraphics[width=.95\columnwidth]{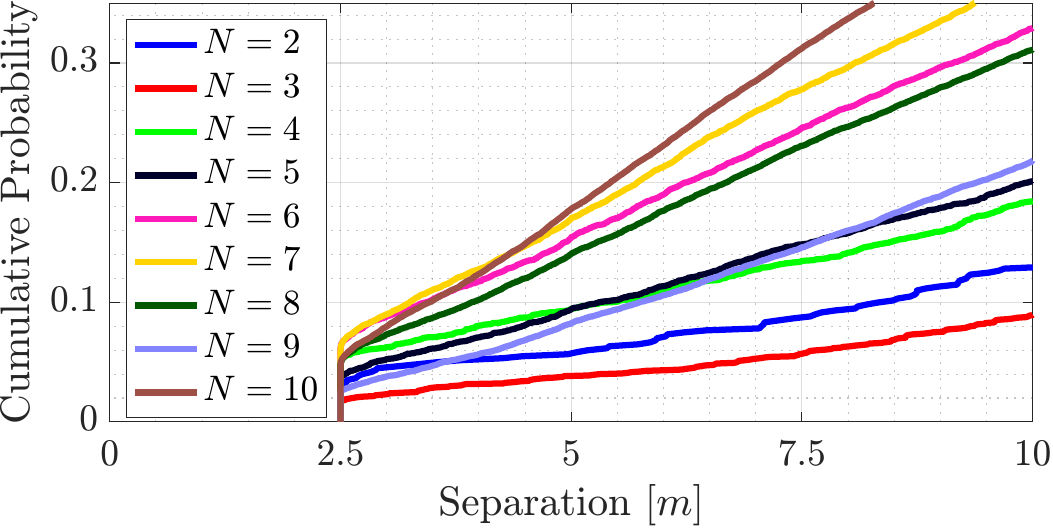}
                \vspace{-11pt}
                \caption{Empirical cumulative distribution functions of separation distance between agents given fleet size $N$. Safety distance was set to $2.5$ meters.}
                \label{fig:collisionGuarantee}
                \vspace{-15pt}
            \end{figure}
            
            To evaluate the proposed method, 300 Monte Carlo simulations were conducted. Agents were randomly placed within a 10-meter radius of the castaways, though castaways might not initially be within any agent's FoV. Simulations considered varying numbers of castaways $\mathcal{C}\in\{3,4,5\}$ and agents $N\in\{1,\ldots,10\}$. Results show significant benefits from the multi-agent approach, as illustrated in Fig. \ref{fig:covVSfleet}. Even with two agents, there is an $87\%$ reduction in average covariance, and a three-agent fleet achieves an additional $10\%$ decrease. However, for missions tracking $5$ targets, adding more UAVs yields minimal or negligible reductions in covariance.

            Inter-agent collision avoidance is crucial for multi-agent systems. The formulation in \ref{ssec:dmpc} ensures agents maintain a minimum separation distance of $d_t=2.5$ meters, as shown in Fig. \ref{fig:collisionGuarantee}. Regardless of fleet size ($N\in\{2,\ldots,10\}$), agents never come closer than the predetermined distance.

            \begin{figure}[t]
       		\centering
    		  \includegraphics[width=.95\columnwidth]{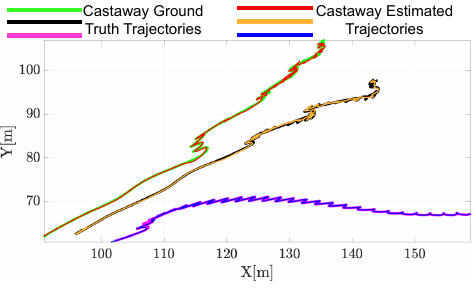}
                \vspace{-13pt}
    		  \caption{Ground truth paths with estimation paths overlay for three targets over the span of 10 minutes.}
    		  \label{fig:estimates}
    		  \vspace{-15pt}
            \end{figure}
            \begin{figure*}
    		  \centering
    		  \includegraphics[width=0.93\textwidth]{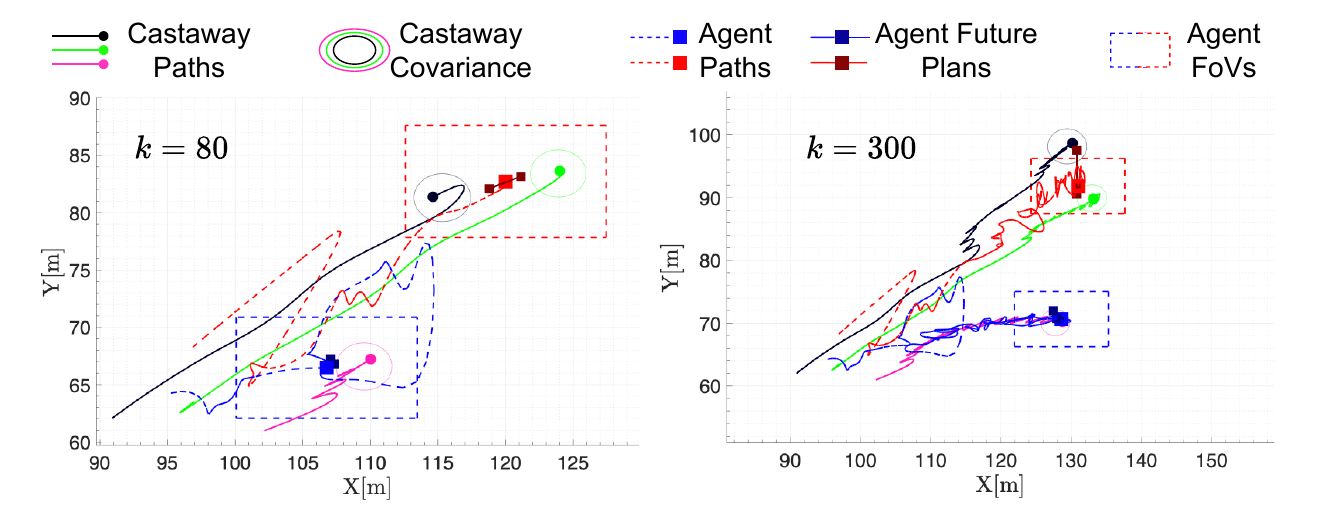}
                \vspace{-15pt}
    		  \caption{Intermediate steps of the proposed multi-agent approach: dotted rectangles indicate the onboard cameras' sensing range, blue and red squares represent agents, and black, green, and magenta circles represent castaways.}
    		  \label{fig:multi-sen}
    		  \vspace{-15pt}
            \end{figure*}
            \begin{figure}
                \centering
                \includegraphics[width=\columnwidth]{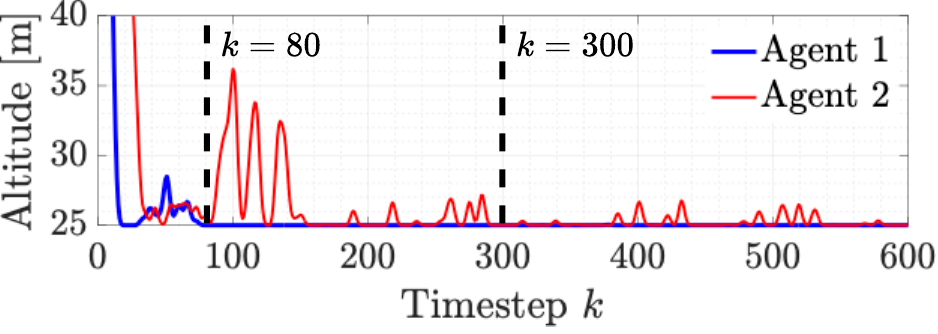}
                \vspace{-22pt}
                \caption{Altitude of the two agents during the simulation. Lowest altitude was set to $25$ meters as a safety measure.}
                \label{fig:altitudes}
                \vspace{-15pt}
            \end{figure}
           
            To analyze the proposed approach, we examine a scenario with $\mathcal{C}=3$ targets and $N=2$ UAV agents. Fig. \ref{fig:estimates} shows the ground truth drift trajectories (black, green, magenta) alongside the estimated trajectories (orange, red, blue), demonstrating accurate estimation with an average root mean square error of $0.18$ meters. In Fig. \ref{fig:multi-sen}, UAV behavior is observed, with targets represented by circles (black, green, magenta) and agents by squares (blue, red). Covariance ellipses and FoV rectangles correspond to each color.

            Snapshots during the first 300 time steps show all targets initially assigned to one group, causing both agents to follow the group and engage in back-and-forth motion for better estimation. After 80 time steps, targets split into two groups: the first agent follows the magenta target, lowering altitude and matching its trajectory, while the second agent observes both remaining targets equally, minimizing their estimation covariance. The second agent adjusts altitude to include both targets in its FoV when possible but primarily lowers altitude for better measurements. Altitude variations are shown in Fig. \ref{fig:altitudes}.
            \vspace{-5pt}
        
    \section{Conclusions} 
    \label{sec:conclusions}
       This research presents an MPC scheme for tracking multiple castaways by calculating UAV control inputs over a receding horizon to minimize position estimation error. Results demonstrate that the method consistently and accurately tracks castaways under realistic conditions, enhancing location precision. Future work will explore adapting the approach to various UAV types and its practical implementation.

    
    {\tiny
    \bibliographystyle{IEEEtran}
    \balance
    \bibliography{bibliography}

\begin{thebibliography}{10}
\providecommand{\url}[1]{#1}
\csname url@samestyle\endcsname
\providecommand{\newblock}{\relax}
\providecommand{\bibinfo}[2]{#2}
\providecommand{\BIBentrySTDinterwordspacing}{\spaceskip=0pt\relax}
\providecommand{\BIBentryALTinterwordstretchfactor}{4}
\providecommand{\BIBentryALTinterwordspacing}{\spaceskip=\fontdimen2\font plus
\BIBentryALTinterwordstretchfactor\fontdimen3\font minus
  \fontdimen4\font\relax}
\providecommand{\BIBforeignlanguage}[2]{{%
\expandafter\ifx\csname l@#1\endcsname\relax
\typeout{** WARNING: IEEEtran.bst: No hyphenation pattern has been}%
\typeout{** loaded for the language `#1'. Using the pattern for}%
\typeout{** the default language instead.}%
\else
\language=\csname l@#1\endcsname
\fi
#2}}
\providecommand{\BIBdecl}{\relax}
\BIBdecl

\bibitem{papaioannou2021towards}
S.~Papaioannou, P.~Kolios, T.~Theocharides, C.~G. Panayiotou, and M.~M.
  Polycarpou, ``Towards automated {3D} search planning for emergency response
  missions,'' \emph{Journal of Intelligent \& Robotic Systems}, vol. 103,
  no.~1, p.~2, 2021.

\bibitem{papaioannou2020coordinated}
S.~Papaioannou, S.~Kim, C.~Laoudias, P.~Kolios, S.~Kim, T.~Theocharides,
  C.~Panayiotou, and M.~Polycarpou, ``Coordinated crlb-based control for
  tracking multiple first responders in 3d environments,'' in \emph{2020
  International Conference on Unmanned Aircraft Systems (ICUAS)}, 2020, pp.
  1475--1484.

\bibitem{anastasiou2020swarm}
A.~Anastasiou, P.~Kolios, C.~Panayiotou, and K.~Papadaki, ``Swarm path planning
  for the deployment of drones in emergency response missions,'' in \emph{2020
  International Conference on Unmanned Aircraft Systems (ICUAS)}.\hskip 1em
  plus 0.5em minus 0.4em\relax IEEE, 2020, pp. 456--465.

\bibitem{anastasiou2021hyperion}
A.~Anastasiou, R.~Makrigiorgis, P.~Kolios, and C.~Panayiotou, ``Hyperion: A
  robust drone-based target tracking system,'' in \emph{2021 International
  Conference on Unmanned Aircraft Systems (ICUAS)}, 2021, pp. 927--933.

\bibitem{anastasiou2023model}
A.~Anastasiou, S.~Papaioannou, P.~Kolios, and C.~G. Panayiotou, ``{Model
  Predictive Control For Multiple Castaway Tracking with an Autonomous Aerial
  Agent},'' in \emph{2023 European Control Conference (ECC)}, 2023, pp. 1--8.

\bibitem{papaioannou2020cooperativeSecurity}
S.~Papaioannou, P.~Kolios, C.~G. Panayiotou, and M.~M. Polycarpou,
  ``Cooperative simultaneous tracking and jamming for disabling a rogue
  drone,'' in \emph{2020 IEEE/RSJ International Conference on Intelligent
  Robots and Systems (IROS)}.\hskip 1em plus 0.5em minus 0.4em\relax IEEE,
  2020, pp. 7919--7926.

\bibitem{papaioannou2021downing}
S.~Papaioannou, P.~Kolios, and G.~Ellinas, ``Downing a rogue drone with a team
  of aerial radio signal jammers,'' in \emph{2021 IEEE/RSJ International
  Conference on Intelligent Robots and Systems (IROS)}, 2021, pp. 2555--2562.

\bibitem{savva2021icarus}
A.~Savva, A.~Zacharia, R.~Makrigiorgis, A.~Anastasiou, C.~Kyrkou, P.~Kolios,
  C.~Panayiotou, and T.~Theocharides, ``{ICARUS}: automatic autonomous power
  infrastructure inspection with {{UAV}}s,'' in \emph{2021 International
  Conference on Unmanned Aircraft Systems (ICUAS)}.\hskip 1em plus 0.5em minus
  0.4em\relax IEEE, 2021, pp. 918--926.

\bibitem{zacharia2023distributed}
A.~Zacharia, S.~Papaioannou, P.~Kolios, and C.~Panayiotou, ``Distributed
  control for {3D} inspection using multi-{UAV} systems,'' in \emph{2023 31st
  Mediterranean Conference on Control and Automation (MED)}.\hskip 1em plus
  0.5em minus 0.4em\relax IEEE, 2023, pp. 164--169.

\bibitem{papaioannou2023integrated}
S.~Papaioannou, P.~Kolios, T.~Theocharides, C.~G. Panayiotou, and M.~M.
  Polycarpou, ``Integrated guidance and gimbal control for coverage planning
  with visibility constraints,'' \emph{IEEE Transactions on Aerospace and
  Electronic Systems}, vol.~59, no.~2, pp. 1276--1291, 2023.

\bibitem{anastasiou2024automated}
A.~Anastasiou, A.~Zacharia, S.~Papaioannou, P.~Kolios, C.~G. Panayiotou, and
  M.~M. Polycarpou, ``{Automated Real-Time Inspection in Indoor and Outdoor 3D
  Environments with Cooperative Aerial Robots},'' in \emph{2024 International
  Conference on Unmanned Aircraft Systems (ICUAS)}, 2024, pp. 496--504.

\bibitem{terzi2019swifters}
M.~Terzi, A.~Anastasiou, P.~Kolios, C.~Panayiotou, and T.~Theocharides,
  ``{SWIFTERS}: A multi-{{UAV}} platform for disaster management,'' in
  \emph{2019 International Conference on Information and Communication
  Technologies for Disaster Management (ICT-DM)}, 2019, pp. 1--7.

\bibitem{papaioannou2024synergising}
S.~Papaioannou, P.~Kolios, C.~G. Panayiotou, and M.~M. Polycarpou,
  ``Synergising human-like responses and machine intelligence for planning in
  disaster response,'' in \emph{2024 International Joint Conference on Neural
  Networks (IJCNN)}, 2024, pp. 1--8.

\bibitem{allianz2023lost}
A.~Group, ``{Safety and Shipping Review 2023},''
  \url{https://commercial.allianz.com/content/dam/onemarketing/commercial/commercial/reports/AGCS-Safety-Shipping-Review-2023.pdf},
  May 2023, accessed on March 11th, 2024.

\bibitem{mediterranean2024migrants}
{UNHCR}, ``Mediterranean situation,''
  \url{https://data2.unhcr.org/en/situations/mediterranean}, 2024, accessed on
  March 4th, 2024.

\bibitem{liu2023survey}
S.~Liu and X.~Li, ``{A Survey on Man Overboard Accident Search and Rescue
  Technology by Unmanned Aerial Vehicle},'' in \emph{2023 35th Chinese Control
  and Decision Conference (CCDC)}, 2023, pp. 1062--1067.

\bibitem{feraru2020Towards}
V.~A. Feraru, R.~E. Andersen, and E.~Boukas, ``{Towards an Autonomous
  {UAV}-based System to Assist Search and Rescue Operations in Man Overboard
  Incidents},'' in \emph{2020 IEEE International Symposium on Safety, Security,
  and Rescue Robotics (SSRR)}, 2020, pp. 57--64.

\bibitem{ramirez2011coordinated}
F.~F. Ram{\'\i}rez, D.~S. Benitez, E.~B. Portas, and J.~A.~L. Orozco,
  ``{Coordinated sea rescue system based on unmanned air vehicles and surface
  vessels},'' in \emph{OCEANS 2011 IEEE-Spain}.\hskip 1em plus 0.5em minus
  0.4em\relax IEEE, 2011, pp. 1--10.

\bibitem{alotaibi2019lsar}
E.~T. Alotaibi, S.~S. Alqefari, and A.~Koubaa, ``{{LSAR}: Multi-{UAV}
  Collaboration for Search and Rescue Missions},'' \emph{IEEE Access}, vol.~7,
  pp. 55\,817--55\,832, 2019.

\bibitem{papaioannou2019decentralized}
S.~Papaioannou, P.~Kolios, T.~Theocharides, C.~G. Panayiotou, and M.~M.
  Polycarpou, ``{Decentralized Search and Track with Multiple Autonomous
  Agents},'' in \emph{2019 IEEE 58th Conference on Decision and Control (CDC)},
  2019, pp. 909--915.

\bibitem{oliveira2016moving}
T.~Oliveira, A.~P. Aguiar, and P.~Encarnação, ``{Moving Path Following for
  Unmanned Aerial Vehicles With Applications to Single and Multiple Target
  Tracking Problems},'' \emph{IEEE Transactions on Robotics}, vol.~32, no.~5,
  pp. 1062--1078, 2016.

\bibitem{song2018multi}
R.~Song, T.~Long, Z.~Wang, Y.~Cao, and G.~Xu, ``{Multi-{UAV} Cooperative Target
  Tracking Method using sparse A search and Standoff tracking algorithms},'' in
  \emph{2018 IEEE CSAA Guidance, Navigation and Control Conference (CGNCC)},
  2018, pp. 1--6.

\bibitem{shen2001theoretical}
H.~H. Shen and Y.~Zhong, ``Theoretical study of drift of small rigid floating
  objects in wave fields,'' \emph{Journal of waterway, port, coastal, and ocean
  engineering}, vol. 127, no.~6, pp. 343--351, 2001.

\bibitem{bishop2001introduction}
G.~Bishop, G.~Welch \emph{et~al.}, ``An introduction to the kalman filter,''
  \emph{Proc of SIGGRAPH, Course}, vol.~8, no. 27599-23175, p.~41, 2001.

\bibitem{anastasiou2023mpc}
A.~Anastasiou, S.~Papaioannou, P.~Kolios, and C.~G. Panayiotou, ``Model
  predictive control for multiple castaway tracking with an autonomous aerial
  agent,'' in \emph{2023 European Control Conference (ECC)}, 2023, pp. 1--8.

\bibitem{richards2007robust}
A.~Richards and J.~P. How, ``Robust distributed model predictive control,''
  \emph{International Journal of control}, vol.~80, no.~9, pp. 1517--1531,
  2007.

\bibitem{julier2017general}
S.~Julier and J.~K. Uhlmann, ``General decentralized data fusion with
  covariance intersection,'' in \emph{Handbook of multisensor data
  fusion}.\hskip 1em plus 0.5em minus 0.4em\relax CRC Press, 2017, pp.
  339--364.

\bibitem{Matzka2009}
\BIBentryALTinterwordspacing
S.~Matzka and R.~Altendorfer, \emph{{A Comparison of Track-to-Track Fusion
  Algorithms for Automotive Sensor Fusion}}.\hskip 1em plus 0.5em minus
  0.4em\relax Berlin, Heidelberg: Springer Berlin Heidelberg, 2009, pp. 69--81.
  [Online]. Available: \url{https://doi.org/10.1007/978-3-540-89859-7_6}
\BIBentrySTDinterwordspacing

\bibitem{sinopoli2004kalman}
B.~Sinopoli, L.~Schenato, M.~Franceschetti, K.~Poolla, M.~I. Jordan, and S.~S.
  Sastry, ``{Kalman filtering with intermittent observations},'' \emph{IEEE
  transactions on Automatic Control}, vol.~49, no.~9, pp. 1453--1464, 2004.

\end{thebibliography}
    }
    
\end{document}